  \documentstyle[prl,aps,multicol,epsf]{revtex}

\draft

\begin{document}

\title{ Lyapunov exponent of many-particle systems:
        testing the stochastic approach  }

\author{ Celia Anteneodo,\thanks{{\rm e-mail: celia@cbpf.br     }}
         Raphael N. P. Maia,\thanks{{\rm e-mail: rapha@cbpf.br     }}
         and Ra\'ul O. Vallejos\thanks{{\rm e-mail: vallejos@cbpf.br  }} }

\address{Centro Brasileiro de Pesquisas F\'{\i}sicas,
         R. Dr. Xavier Sigaud 150, \\
         22290-180, Rio de Janeiro, Brazil}

\date{\today}

\maketitle

\begin{abstract}

The stochastic approach to the determination of the largest Lyapunov 
exponent of a many-particle system is tested in the so-called mean-field 
XY-Hamiltonians. 
In weakly chaotic regimes, the stochastic approach relates 
the Lyapunov exponent to a few statistical properties of the Hessian matrix of the
interaction, which can be calculated as suitable thermal averages.
We have verified that there is a satisfactory quantitative agreement between 
theory and simulations in the disordered phases of the XY models, either
with attractive or repulsive interactions. 
Part of the success of the theory 
is due to the possibility of predicting the shape of the required correlation 
functions, because this permits the calculation of correlation times as
thermal averages.

\end{abstract}



 \begin{multicols}{2}

\narrowtext

\section{Introduction}
\label{sec:introduction}

In a recent paper \cite{vallejos02} we presented a theory 
(``the stochastic approach") that allows to estimate the largest Lyapunov 
exponent of many-particle systems having 
a smooth Hamiltonian. This theory is inspired by and complementary to the 
work of Pettini et al. \cite{geometric,casetti96}, and Barnett et al. \cite{barnett96}.

Starting from the linear equations that describe the evolution of tangent vectors, 
we used in \cite{vallejos02} the techniques of stochastic linear differential equations to 
obtain an average propagator whose diagonalization provides the desired Lyapunov 
exponent. Even though the result obtained in this way is formally (almost) exact,
approximations must be invoked in practical situations to calculate a concrete Lyapunov exponent.  

The aim of this paper is to show that, even if one makes several crude approximations,
the theory still keeps the strength to describe {\em quantitatively} some 
weakly chaotic many-particle systems. We do this by comparing the predictions
of the theory with the outcomings of numerical simulations of the Hamiltonian 
many-particle dynamics.

The paper is organized in the following way. 
For reasons of self-containedness, we begin by presenting a short review of the 
theory in Sect.~\ref{sec:theory}. 
The systems to be studied (the mean-field XY-Hamiltonians) are 
introduced in Sect.~\ref{sec:XY}, where we also work out the predictions of
the theory for these particular systems. 
Section~\ref{sec:num} contains the critical comparison 
of theoretical results with numerical simulations.
%
%
We close in Sect.~\ref{sec:conclusions} with some remarks.

 \section{Review of the theory}
 \label{sec:theory}

The theory developed in \cite{vallejos02} can, in principle, be applied 
to any dynamical system. However, given that the approach is perturbative, 
its success will in general depend on the choice of the unperturbed system. 
Here we restrict ourselves to perturbations of ballistic motion, i.e., 
Hamiltonians of the type
\begin{equation}
\label{ham}
{\cal H} = \frac{1}{2I} \sum_{i=1}^N p^2_i + {\cal V}(q_1,\ldots,q_N),
\end{equation}
with $q_i$ and $p_i$ conjugate position-momentum coordinates, $I$ the 
mass (or moment of inertia)
of the particles, and the interaction ${\cal V}$ that 
we assume to be small.

Let us denote phase space points by
$x= (q_1,\ldots,q_N, p_1,\ldots, p_N)$  
and tangent vectors by
$\xi= (\delta q_1,\ldots,\delta q_N, \delta p_1,\ldots, \delta p_N)^T$. 
Differentiating the Hamilton equations, one obtains the evolution 
equations for tangent vectors:  
\begin{equation}
\label{tangent0}
\dot \xi =   {\bf A}(t)\,\xi.
\end{equation}
For a Hamiltonian of the special form (\ref{ham}), 
the operator $\bf A$ has the simple structure
\begin{equation} \label{A}
{\bf A}(t) =  
\left( \matrix{    0          & \openone/I \cr
                -{\bf V}(t)   &    0          }\right) \; .
\end{equation}
Here $\bf V$ is the Hessian matrix of the potential ${\cal V}$, namely
\begin{equation} \label{V}
V_{ij} \;=  \;  \frac{\partial^2{\cal V}}{\partial q_i \partial q_j} \;.
\end{equation}
Once initial conditions $x_0$ and $\xi_0$ have been specified, the
Lyapunov exponent $\lambda$ is found by calculating the 
limit\cite{benettin76}
\begin{equation}
\label{defliapunov}
\lambda =   \lim_{t \to \infty} \frac{1}{2t} \ln | \xi (t; x_0,\xi_0)|^2 \; .
\end{equation}
We assume that for any initial condition $x_0$ in phase space,
the trajectory $x(t;x_0)$ is ergodic on its energy shell.
This implies that $\lambda$ depends only on energy and other system
parameters, but not on $x_0$, which can then be chosen randomly according to 
the microcanonical distribution. There will also be no dependence on initial 
tangent vectors, because if $\xi_0$ is also chosen randomly, it will have a 
non-zero component along the most expanding direction. 
Moreover, if the corrections to the exponential law implicit in 
Eq.~(\ref{defliapunov}) are neglected, one can write
\begin{equation}
\label{defliapunov2}
\lambda \approx   \lim_{t \to \infty} \frac{1}{2t} \ln 
\left\langle | \xi (t; x_0,\xi_0)|^2 \right\rangle\; ,
\end{equation}
brackets meaning microcanonical averages over $x_0$.
This approximation will be analyzed in Sect.~\ref{sec:num}
with the help of numerical simulations.

By letting $x_0$ be a random variable, ${\bf V}(t;x_0)$
becomes a stochastic process, and Eq.~(\ref{tangent0})
can be treated as a stochastic differential equation.
However, as we are interested in the square of the norm of $\xi$, 
we focus, not on Eq.~(\ref{tangent0}) itself, but in the related
equation for the evolution of the ``density matrix" $\xi \xi^T$:
\begin{equation} \label{vK0}
\frac{\rm d}{{\rm d} t} (\xi\xi^T) =   {\bf A}\xi\xi^T  +
\xi\xi^T{\bf A}^T \equiv
\widehat{\bf A} \xi\xi^T,
\end{equation}
the rightmost identity defining the linear superoperator $\widehat{\bf A}$.
For the purpose of the perturbative approximations to be done, the
operator $\widehat{\bf A}$ is split into two parts
\begin{equation}
\widehat{\bf A} =  \widehat{\bf A}_0 + \widehat{\bf A}_1(t) \; ,
\end{equation}
where $\widehat{\bf A}_0$ corresponds to the evolution in the absence of
interactions. In our case $\widehat{\bf A}_0$ and $\widehat{\bf A}_1$
are associated with
\begin{equation} \label{a0a1}
{\bf A}_0 =         \left( \matrix{  0      &  \openone \cr
                                     0      &  0             }\right)
\qquad \mbox{and} \qquad
{\bf A}_1 =    \left( \matrix{       0      &  0        \cr
                                -{\bf V}(t) &  0             }\right) \, ,
\end{equation}
respectively (we have set $I=1$, but it can be brought back at any moment by 
dimensional considerations).
Whenever ${\bf A}_1(t)$ is small, it is possible to manipulate Eq.~(\ref{vK0}) to 
derive an explicit expression for the evolution of the {\em average} of 
$\xi\xi^T$:
\begin{equation}
\label{solution}
  \langle {\xi\xi^T}\rangle(t) = e^{ t\widehat{\bf \Lambda}} \; {\xi_0\xi_0^T} \; ,
\end{equation}
where $\widehat{\bf \Lambda}$ is a time-independent superoperator given 
by the perturbative expansion:
\begin{eqnarray} \label{expansion} \nonumber
& & \widehat{\bf \Lambda}  =   
   \widehat{\bf A}_0 + \langle \widehat{\bf A}_1 \rangle \\ \label{final}
& & +
\int_0^\infty {\rm d} \tau
\left\langle\delta \widehat{\bf A}_1(t)      \, e^{ \tau \widehat{\bf A}_0 } \,
             \delta\widehat{\bf A}_1(t-\tau) \, e^{-\tau \widehat{\bf A}_0 }
\right\rangle + \cdots \; ,
\end{eqnarray}
with
\begin{equation}
\delta \widehat{\bf A}_1(t)= \widehat{\bf A}_1(t) -\langle \widehat{\bf A}_1
\rangle \; .
\end{equation}
A clear exposition of this derivation, together with a detailed
discussion of its domain of validity has been given by van Kampen 
\cite{vankampen}.

Let $L_{\rm max}$ be the  eigenvalue of $\widehat{\bf \Lambda}$
which has the largest real part. Taking the trace of Eq.~(\ref{solution}), one sees 
that the largest Lyapunov exponent $\lambda$ is 
related to the real part of $L_{\rm  max}$ by
\begin{equation}
\lambda =   \mbox{$\frac{1}{2}$} \, \mbox{Re} \, \left( L_{\rm max} \right) \;.
\end{equation}

In Eq.~(\ref{expansion}) we give explicitly only the first two 
cumulants, the dots stand for third cumulants and higher order ones. 
If all cumulants were summed up, Eq.~(\ref{solution}) would be exact 
in the long-time regime $t \gg \tau_c$ \cite{vankampen}.
From now on, we restrict our analysis to the propagator 
$\widehat{\bf \Lambda}$ truncated at the second order.
The perturbative parameter controlling the quality of the truncation 
is the product of two quantities, the ``Kubo number" $\sigma \tau_c$.
The first factor, $\sigma$, characterizes the amplitude of the 
fluctuations of $\widehat{\bf A}_1(t)$.
The second, $\tau_c$, is the correlation time of $\widehat{\bf A}_1(t)$.

%
%

To proceed further one needs the matrix of $\widehat{\bf \Lambda}$ 
in some basis. The crudest approximation consists in restricting 
$\widehat{\bf \Lambda}$ to the subspace spanned by the following three 
matrices:
\begin{equation} \label{i1i2i3}
{\bf I}_1=  \left( \matrix{  \openone  &       0    \cr
                               0       &       0     } \right), \;
{\bf I}_2=  \left( \matrix{    0       &       0    \cr
                               0       &   \openone  } \right), \;
{\bf I}_3=  \left( \matrix{    0       &   \openone \cr
                            \openone   &       0     } \right).
\end{equation}
This choice is equivalent to a {\em mean-field approximation in 
tangent space} \cite{vallejos02}.
After some lengthy algebra one arrives at the corresponding 3$\times$3 
matrix for $\widehat{\bf \Lambda}$:
\begin{equation}
\label{iso}
{\bf \Lambda} =  
\left( \matrix{ 0                       &  0                      &  2                      \cr
                2\sigma^2 \tau_c^{(1)}  & -2\sigma^2 \tau_c^{(3)} & -2\mu                   \cr
          -\mu+ 2\sigma^2 \tau_c^{(2)}  &  1                      & -2\sigma^2 \tau_c^{(3)} \cr
  } \right)                                \; ,
\end{equation}
with the definitions
\begin{eqnarray}
\mu            &=& \frac{1}{N} {\rm Tr } \langle {\bf V} \rangle                          \; ,  \label{mu} \\
\sigma^2       &=& \frac{1}{N} {\rm Tr } \langle \left( \delta {\bf V} \right)^2 \rangle  \; ,  \label{s2} \\
\tau_c^{(k+1)} &=& \int_0^{\infty} d\tau \, \tau^k f(\tau)                                      \label{tau} \; ,
\end{eqnarray}
where we have introduced the normalized correlation function $f(\tau)$
\begin{equation}
\label{corr}
f(\tau) = \frac{1}{N \sigma^2} {\rm Tr } \langle \delta {\bf V}(0)
                                                 \delta {\bf V}(\tau) \rangle   \;.
\end{equation}
[A Lyapunov exponent with the correct units (inverse time) is obtained
 by making the substitutions $\mu, \sigma \to \mu/I, \sigma/I$ in Eq.~(\ref{iso}).]

In the mean-field approximation the Lyapunov exponent is expressed in 
terms of the set of four parameters
$\mu$ and $\sigma^2 \tau_c^{(k+1)}$, $k=0,1,2$.
The parameters $\mu$ and $\sigma$ are, respectively, the mean and 
variance of the stochastic process ${\bf V}(t)$, and can in principle be 
obtained analytically by calculating the corresponding microcanonical averages.

The characteristic time $\tau_c^{(1)} \equiv \tau_c $ is naturally interpreted 
as the correlation time of the process ${\bf V}(t)$. 
For instance, if $f(\tau)$ is approximately Gaussian, 
the expansion of
$ \langle  {\bf V}(0) {\bf V}(\tau) \rangle $
around $\tau=0$ gives an explicit formula for the correlation
time, namely
\begin{equation}
\frac{1}{\tau_c}=  
\label{tauc1}
\left[ \frac{2}{\pi \sigma^2 N}
\mbox{Tr} \left \langle \left( \frac{d{\bf V}}{dt} \right)^2 \right 
\rangle
\right]^{1/2} \; .
\end{equation}
In this case $\tau_c^{(2)}$ and $\tau_c^{(3)}$ are trivially related to $\tau_c$:
\begin{eqnarray}
\tau_c^{(2)} & =  & \frac{2}{\pi}  \tau_c^2 \; ,   \\
\tau_c^{(3)} & =  & \frac{2}{\pi}  \tau_c^3 \; .
\end{eqnarray}

\section{The mean-field XY-Hamiltonian}
\label{sec:XY}

In this section we start the application of the perturbative/mean-field
theory of Sect.~\ref{sec:theory} to a specific model. 
Consider the one-dimensional Hamiltonian
\begin{equation}
H   =  \frac{1}{2I} \sum_{i=1  }^N     L_{i}^{2} +
       \frac{J}{2N} \sum_{i,j=1}^N 
       \left[   1-\cos(\theta_{i}-\theta_{j}) \right] \; .
\label{HXY}
\end{equation}
This is the so-called mean-field XY-Hamiltonian.
It represents a lattice of classical spins with infinite-range interactions.
Each spin rotates in a plane and is therefore described by an angle 
$0 \le \theta_i < 2\pi$, and its conjugate angular 
momentum $L_i$, with $i=1,\ldots,N$. 
The constants $I$ and $J$ are the moment of inertia and the interaction strength, 
respectively. (Of course, one can also think of point particles of mass $I$ moving
on a circle.)

The mean-field XY-Hamiltonian (HMF) has been extensively studied in the 
last few years (see \cite{reviewHMF} for a review). 
The reasons for the interest in this model are various. 
From a general point of view, the HMF can be considered the simplest 
prototype for complex systems with long-range interactions like galaxies and plasmas 
(in fact, the HMF is a descendant of the mass-sheet gravitational model 
\cite{antoni95}). But the HMF is also interesting for its anomalies, 
be them model-specific or not.
Especially worth of mention are the long-lived quasi-equilibrium 
states observed in the ferromagnetic HMF. 
These states exhibit breakdown of ergodicity, anomalous diffusion, and non-Maxwell 
velocity distributions \cite{metastable}. 
The explanation of these unusual behaviors may require an extension of the 
standard statistical mechanics, e.g., along the lines proposed by Tsallis \cite{tsallis}.
(Interesting anomalies are also present in the anti-ferromagnetic HMF \cite{bicluster}).

The simplicity of the HMF makes possible a full analysis of its statistical
properties either in the canonical \cite{antoni95} or microcanonical 
ensemble \cite{antoni02,velazquez03}.
If interactions are attractive ($J>0$), the system exhibits a ferromagnetic 
transition at the critical energy $E_c=3JN/4$. 
In the case $J<0$ there is no ordered phase with finite magnetization, and,
for not too low temperatures, the system behaves like the disordered phase of the 
$J>0$ case. However, at small energies, some kind of order 
appears, leading to a complex dynamical behavior \cite{bicluster}.

Lyapunov exponents have been studied with detail, both numerically and analytically. 
For the ferromagnetic case, the simulations \cite{reviewHMF,latora99,yamaguchi96}
show that in the magnetized phase the Lyapunov exponent remains finite in the 
thermodynamic limit $N \to \infty$. In contrast, if $E>E_c$, $\lambda$ goes to zero when 
$N \to \infty$; the same behavior is also observed in the antiferromagnetic case
in all the energy range \cite{reviewHMF,latora00}. 
%
%
The first theoretical studies were due to Firpo \cite{firpo98}, who derived analytical
expressions using the geometric method \cite{geometric,casetti96} 
(see also the discussion in Ref.~\cite{latora00}). 
Scaling laws in the high-energy regimes were derived as well using a random-matrix 
approach \cite{anteneodo01}.
At low temperatures the predictions
of the geometric method for the HMF's Lyapunov exponent are not satisfactory.
There is solid evidence that the geometric method predicts
wrong scaling laws at low temperatures for both ferromagnetic and antiferromagnetic
interactions \cite{latora99,latora00}. 
This means that there are important gaps in the 
theoretical description
of the Lyapunov exponent of the HMF (and of many-particle systems, in general). 
Further studies are necessary for understanding the precise domain of validity of
the existing theories: this knowledge will be used to make the
corresponding improvements.

Going back to the stochastic approach, it was proven in \cite{vallejos02} that the 
mean-field approximation is exact in the HMF (this agrees with the supersymmetric 
analysis of T\u{a}nase-Nicola and Kurchan \cite{tanase02}).
In the disordered phases of the HMF the fluctuations are small (see below), so, it is expected 
that the second-order perturbative approximation will work well, irrespective of the sign of $J$.

In order to test the stochastic approach, in its mean-field second-order-perturbative version,
one has to calculate the average, variance, and correlation function of the Hessian ${\bf V}(t)$, 
i.e., Eqs.~(\ref{mu}), (\ref{s2}) and (\ref{corr}). These ingredients are microcanonical averages
of the appropriate observables. In the disordered regimes we will consider, microcanonical 
and canonical averages are equivalent (to leading order in $N$), so, we will prefer the simpler 
canonical
averaging. Anyway, we will verify numerically that canonical and time averages coincide.

\subsection{Calculation of $\mu$}

Before embarking in the calculation of canonical averages, it is
convenient to write the Hamiltonian (\ref{HXY}) in the simplified form:
\begin{equation}
\label{hamprime}
H   =    \frac{1}{2I} \sum_{i=  1  }^N L_{i}^{2} + \frac{JN}{2} \left( 
1-m^2 \right) \; ,
\end{equation}
where we have introduced the magnetization per particle 
\begin{equation}
{\bf m} =   \frac{1}{N} \sum_{i=  1}^N \hat{\bf r}_i  \; ,
\end{equation}
with
\begin{equation}
\hat{\bf r}_i =   (\cos \theta_i,\sin \theta_i) \; .
\end{equation}
In terms of ${\bf m}$ and $\{\hat{\bf r}_i\}$ the elements of the 
Hessian matrix 
read
\begin{eqnarray}
V_{ii} & =   & J( {\bf m} \cdot \hat{\bf r}_i - \frac{1}{N} )  \; , \\
V_{ij} & =   & -J \hat{\bf r}_i \cdot \hat{\bf r}_j  \;, \qquad i \ne j   \; . 
\end{eqnarray}
Then one has
\begin{equation}
\label{muoverj}
\mu/J=   \left\langle m^2 \right\rangle -\frac{1}{N}  \; .
\end{equation}
Our next objective is the average $\left\langle m^2 \right\rangle$.
The interacting part of the canonical partition function reads 
\begin{equation}
Z(\beta) \propto \int d{\bf \theta} \, {\mbox e}^{\beta J N m^2 /2}  \; .
\end{equation}
The integration over the angles can be reduced
to an integration over the possible magnetizations:
\begin{equation}
\label{partition}
Z(\beta) \propto \int d{\bf m} \, g({\bf m}) \, {\mbox e}^{\beta J N m^2 /2} \; .
\end{equation}
Here $g({\bf m})$ represents the density of states in $\theta$-space 
with magnetization ${\bf m}$. 
The problem of finding $g({\bf m})$ has a long history and is known
as Pearson's random-walk problem \cite{hughes95}. 
There are no simple closed expressions for $g({\bf m})$, but 
as we are interested in the disordered phases, where $m$ is of the order 
of $1/\sqrt{N}$, it suffices to use the central-limit approximation: 
\begin{equation}
\label{centrallimit}
g({\bf m}) \propto \mbox{e}^{-Nm^2} \; .
\end{equation}
The relative error of this approximation is \cite{hughes95} 
\begin{equation}
1 - \frac{1}{2N}\left( 1-2Nm^2+\frac{1}{2}N^2m^4 \right)+ \ldots   \; .
\end{equation}
So, we can safely use expression (\ref{centrallimit}) if 
$N \gg 1$ and the relevant configurations are such that $Nm^2 \sim 1$. 
The last condition is synonymous with disorder. 
It is always satisfied in the antiferromagnetic case. 
In the case $J>0$ it is satisfied if the energy is above the critical value 
$\varepsilon = 3J/4$, but not very close to it. 

Putting together the formulas (\ref{partition}) and (\ref{centrallimit}) we get 
the probability distribution of ${\bf m}$. To leading order in $1/N$ we have
\begin{equation}
P({\bf m}) \propto \exp \left[ -N \left(1-\frac{\beta J}{2}\right)m^2  
\right] \; .
\end{equation}
And then we arrive at
\begin{equation}
\label{m2}
\left\langle m^2 \right\rangle \approx   \frac{1}{N \left( 1-\beta J/2 \right)}  \; 
\end{equation}
(here and in the following, $\approx$ means ``equal to leading order
in $1/N$"). 
This expression, together with (\ref{hamprime}) and the equipartition theorem, 
$kT=\langle L_i^2 /I \rangle$,
allows us to obtain the relationship between the temperature $T$ and 
the energy per particle $\varepsilon= E/N$:
\begin{equation}
kT \approx 2 \varepsilon - J \; .
\end{equation}
As a function of $\varepsilon$, the average $\mu$ reads
\begin{equation}
\mu/J \approx \frac{1}{N \left( 4\varepsilon/J - 3 \right)} \; .
\label{muoverjfinal}
\end{equation}
Combining Eqs.~(\ref{muoverj}) and (\ref{m2}) one sees that $\mu$ is a 
finite-temperature correction to the $T=\infty$ magnetization.
So, for fixed $N$, $\mu$ goes to zero as the energy is increased.
Equation~(\ref{muoverjfinal}) coincides with the result obtained by
Firpo using a different technique \cite{firpo98}.

\subsection{Calculation of $\sigma^2$}
\label{sec:sigma2}

Given that all degrees of freedom are statistically equivalent, the 
definition of $\sigma^2$, Eq.~(\ref{s2}), can be expressed as
\begin{equation}
\sigma^2 =   \left\langle \left[ \left( \delta {\bf V} \right)^2\right]_{11} \right\rangle 
         =   \sum_{j=  1}^N \left\langle \left( \delta V_{1j} \right)^2 \right\rangle 
\; ,
\end{equation}
the rightmost identity following from the symmetry of ${\bf V}$. 
The abovementioned equivalence can be used once more, together with the 
definition of $V_{ij}$, to obtain
\begin{eqnarray}
\frac{\sigma^2}{J^2} & =   & 
\left\langle \left( {\bf m} \cdot \hat{\bf r}_1 \right)^2 \right\rangle 
-
\left\langle {\bf m} \cdot \hat{\bf r}_1  \right\rangle^2 \nonumber \\ & 
& +
\frac{1}{N} \left[   
\left\langle \left( \hat{\bf r}_1 \cdot \hat{\bf r}_2 \right)^2 
\right\rangle -
\left\langle \hat{\bf r}_1 \cdot \hat{\bf r}_2 \right\rangle^2   \right] 
\; .
\label{s2prime}
\end{eqnarray}
To proceed with the evaluation of $\sigma^2$ one needs the probability 
distributions for two and three particles, $P_2(\theta_1,\theta_2)$ and 
$P_3(\theta_1,\theta_2,\theta_3)$. Consider first $P_2$, which is just
\begin{equation}
P_2(\theta_1,\theta_2) \propto \int d\theta_3 \ldots d\theta_N \, 
{\mbox e}^{\beta J N m^2 /2}  \; .
\end{equation}
To do the integrations we split the magnetization into two parts: 
the contribution from particles 1 and 2, and the remainder:
\begin{equation}
{\bf m} = {\bf m}_{12} + {\bf m}_{3N} \; ,
\end{equation}
where
\begin{eqnarray}
{\bf m}_{12} & = & \frac{1}{N} \left( \hat{\bf r}_1 + \hat{\bf r}_2 \right) \; ,     \\
{\bf m}_{3N} & = & \frac{1}{N} \sum_{i=3}^N  \hat{\bf r}_i       \; .
\end{eqnarray}
We outline the final steps. 
Like done before, we introduce a density of configurations $g({\bf m}_{3N})$:
\begin{equation}
P_2(\theta_1,\theta_2) \propto \int \, d{\bf m}_{3N} \; g({\bf m}_{3N})
\exp \left[ \frac{\beta J N}{2} ({\bf m}_{12}+{\bf m}_{3N})^2 \right] \; .
\end{equation}
Now invoke the central limit theorem to approximate $g({\bf m}_{3N})$.
Switch to polar coordinates. Integrate over the polar angle, and then
over the modulus $m_{3N}$ (the upper limit of integration can be extended
to infinity). The final result is
\begin{equation}
\label{p2}
P_2(\theta_1,\theta_2) \approx A \; 
\exp \left[ \, \alpha  \left( \hat{\bf r}_1 \cdot \hat{\bf r}_2  \right) \right]  \; ,
\end{equation}
where $A$ is a normalization constant and 
\begin{equation}
\alpha =  \frac{2}{N \left( 4\varepsilon/J - 3 \right)}   \; .
\end{equation}
The parameter $\alpha$ is the relevant perturbative quantity in this problem.
It is small whenever $N \gg 1$ {\em and} the energy exceeds the transition 
value by a finite, large enough, amount. (Notice that $\alpha = 2\mu/J$.)

Proceeding in the same way one also finds the three-particle distribution 
function:
\begin{equation}
P_3(\theta_1,\theta_2,\theta_3) \approx B \; 
\exp \left[ \, \alpha  \left( \hat{\bf r}_1 \cdot \hat{\bf r}_2 + 
                              \hat{\bf r}_2 \cdot \hat{\bf r}_3 +
                              \hat{\bf r}_3 \cdot \hat{\bf r}_1    \right) \right]  \; .
\end{equation}
With the distribution functions $P_2$ and $P_3$ in our hands, we go back
to the calculation of $\sigma^2$, Eq.~(\ref{s2prime}). The moments of 
$\hat{\bf r}_1 \cdot \hat{\bf r}_2$ are immediate:
\begin{eqnarray}
\langle \hat{\bf r}_1 \cdot \hat{\bf r}_2 \rangle 
& \approx & \frac{\alpha}{2}   \; , \\
\left \langle \left( \hat{\bf r}_1 \cdot \hat{\bf r}_2 \right)^2 \right \rangle 
& \approx & \frac{1}{2}        \; .
\end{eqnarray}
In addition one has
\begin{equation}
\langle {\bf m} \cdot \hat{\bf r}_1 \rangle \approx 
\frac{1}{N} + \langle \hat{\bf r}_1 \cdot \hat{\bf r}_2 \rangle = {\cal O} (1/N) \; .
\end{equation}
Then it is easy to verify that Eq.~(\ref{s2prime}) becomes
\begin{equation}
\frac{\sigma^2}{J^2} \approx  
\left\langle \left( {\bf m} \cdot \hat{\bf r}_1 \right)^2 \right\rangle  + \frac{1}{2N} \; .
\end{equation}
Splitting ${\bf m}$ into its $N$ parts, and using that particles are statistically equivalent, 
we arrive at
\begin{equation}
\frac{\sigma^2}{J^2} \approx  \frac{1}{N} + 
\langle \left( \hat{\bf r}_1 \cdot \hat{\bf r}_2 \right) 
        \left( \hat{\bf r}_1 \cdot \hat{\bf r}_3 \right) \rangle \; .
\end{equation}
The three-body average is quickly done (with the help of {\em Mathematica} \cite{wolfram91})
by expanding $P_3$ in powers of $\alpha$:
\begin{equation}
\langle \left( \hat{\bf r}_1 \cdot \hat{\bf r}_2 \right) 
        \left( \hat{\bf r}_1 \cdot \hat{\bf r}_3 \right) \rangle 
\approx \frac{\alpha}{4} \; .
\end{equation}
So, the final expression for $\sigma^2$ is 
\begin{equation}
\sigma^2 \approx \frac{J^2}{N}  \left( 1 + \frac{N\alpha}{4} \right) \; .
\label{sigma2final}
\end{equation}
%

\subsection{Correlation function and correlation time}

Above the transition, the relative importance of the interactions decreases 
with increasing $\varepsilon$ and $N$, and the dynamics is dominated by the 
kinetic part of the Hamiltonian.
The picture is that of particles rotating almost freely during times which 
are long as compared to the mean rotation period. If the system is in 
equilibrium, the dynamics can be modeled by the free-motion equations 
\begin{equation}
\theta_k \approx \theta_k (t=0) + L_k t/I \; ,  \qquad 1 \le k \le N \; ,
\end{equation}
where $\{ \theta_k(t=0) \}$ and $\{L_k\}$ are independent random 
variables with uniform and Maxwell distributions, respectively. 
This is a first approximation valid only during short times.
Then it is easy to show that the correlation functions of the elements
of the Hessian are directly related to the characteristic function of
the momentum distribution, i.e.,
\begin{equation}
2 \left\langle 
\cos \left( \theta_k - \theta_j \right)(\tau) 
\cos \left( \theta_k - \theta_j \right)(0) 
\right\rangle 
\approx   
\left| \left\langle \exp \left( -i L_k \tau /I\right) \right\rangle \right|^2 \; .
\end{equation}
In equilibrium the characteristic function is Gaussian:
\begin{equation}
\left| \left\langle \exp \left( -i L_k \tau /I \right) \right\rangle \right|^2 = 
\exp \left[ -\left( \tau/\tau_* \right)^2 \right] \; ,
\end{equation}
with 
\begin{equation}
\tau_* = \sqrt{\frac{I}{kT}} \; .
\end{equation}
After using the definition (\ref{tau}) we arrive at the simplest estimate of 
the correlation time in the high-energy regime
\begin{equation}
\label{tauc2}
\tau_c \sim \sqrt{\frac{\pi I}{4kT}} \; .
\end{equation}
The correlation time $\tau_c$ is of the order of the mean period of rotation.
It is independent of the system size because it is not directly 
associated with interactions. Of course, interactions are responsible for the 
Maxwell equilibrium distributions, and for the extinction of the ballistic regime
and its substitution by a random-walk one. However, the time scales involved in these 
processes are much longer than $\tau_c$: the first one is the relaxation time for 
the one-body momentum distribution, the second is the momentum correlation time. 
Both times grow with $N$, and are not related with $\tau_c$.

A more precise estimate for $\tau_c$ can be derived if the correlation function is 
indeed Gaussian. Then the correlation time is that given by  Eq.~(\ref{tauc1}). 
We will not detail the calculation of the canonical averages of Eq.~(\ref{tauc1})
because they are very similar to those of Sect.~\ref{sec:sigma2}. We just show the
result:
\begin{equation}
\label{taucfinal}
\tau_c \approx  
\sqrt{ \frac{\pi I}{4kT} 
       \left(  \frac{ 1 + N \alpha /4 }{1 + N \alpha /8 } \right) } \; ,
\end{equation}
very close to the simple estimate of Eq.~(\ref{tauc2}).

\subsection{The Lyapunov exponent}

Gathering the results of previous sections, one can construct the $3 \times 3$ 
matrix ${\bf \Lambda}$ associated to the average propagator for the HMF. 
The Lyapunov exponent is extracted from the eigenvalue of ${\bf \Lambda}$ with the 
largest real part. Though the general expression for $\lambda$ might be written 
down explicitly, its content would not justify its extension. 
Notwithstanding, it is worth exhibiting the leading term in $1/N$.
Notice that both $\mu$ and $\sigma^2$ are of order $1/N$, and that 
$\tau_c^{(k)} = {\cal O}(1)$ (once the correlation function 
is assumed Gaussian). Then one gets
\begin{equation}
\label{lambda-asymp}
\lambda \approx  \left(  \frac{\sigma^2 \tau_c}{2} \right)^{1/3} + 
{\cal O}\left( N^{-2/3} \right) = {\cal O}\left( N^{-1/3} \right) \; ,
\end{equation}
with $\sigma^2$ and $\tau_c$ given by Eqs.~(\ref{sigma2final}) and (\ref{taucfinal}).
The absence of $\mu$ in the leading-order expression for $\lambda$ is a reflection of the fact that in the
disordered phases of the HMF, fluctuations are much larger than the average, i.e. $\sigma \gg \mu$,
and dominate the tangent dynamics.

\section{Numerical studies} 
\label{sec:num}

For comparing the theory with simulations, we will use the data existing
in the literature \cite{latora00,latora98,anteneodo98} as well as some
additional data generated by us. We give a succinct description of how our 
simulations were made.
Hamilton's equations of the $N$-particle system were evolved using the Neri-Yoshida 
fourth-order symplectic algorithm \cite{yoshida90}. The time step was fixed through all 
simulations to $\Delta t=0.1$ (units are such that $I=1$ and $|J|=1$).
Initial conditions were chosen randomly: angles with a uniform
distribution in $[0,2\pi]$, and angular momenta Gaussian distributed. 
Then all momenta were shifted by a fixed amount to set to zero the total 
momentum (a constant of motion). Finally, velocities were multiplied by an
appropriate factor to fix the total energy at a chosen value $N\varepsilon$.
The initial one-body distributions generated in this way are close to their equilibrium 
values in the disordered regimes defined by $\varepsilon > 3J/4$ and $J>0$, or 
$\varepsilon > 0$ and $J<0$.
Anyway, before doing any ``measurements" the systems were allowed to relax during
a time we call $t_{\rm eq}$, typically $t_{\rm eq} \in [1000,10000]$. 
Concerning Lyapunov exponents, we used Benettin's standard algorithm \cite{benettin76}. 
The initial Euclidian distance between a trajectory and its
companion were set to $d_0=10^{-6}$. We used two alternative renormalization
procedures: (a) The distance vector was compressed each
time the distance exceeded $d_0 f$; usually with $f=10$ or $f=20$;
(b) Renormalizations were made at equally spaced instants, the interval between
successive renormalizations corresponding, in average, to expansion factors $f \approx 10,20$.

Numerically, the Lyapunov exponent is estimated by averaging over initial conditions
which are propagated during a finite time:
\begin{equation}
\label{liapunovnum}
\lambda(t) \equiv  \frac{1}{2t} \left \langle \ln | \xi (t; x_0,\xi_0)|^2 \right \rangle \; .
\end{equation}
We usually considered 10 pairs of randomly-chosen initial conditions. 
Each pair was propagated until a time $t=t_{\rm prop}$, when we judged 
that $(1/t)\ln |\xi|$ had converged to a limiting value.
Typically, $t_{\rm prop} \in [1000,5000]$. Remember, however, that our theoretical scheme commutes logarithm 
and average. In order to test this approximation, we also computed the average
\begin{equation}
\label{liapunovnum2}
\lambda^\star(t) \equiv   \frac{1}{2t} \ln \left\langle | \xi (t; x_0,\xi_0)|^2 \right\rangle\; .
\end{equation}

In many cases we run tests to verify that our numerical results are robust against 
suitable changes of initial conditions, renormalization procedure, 
or the set of parameters $\{ t_{\rm eq}, t_{\rm prop}, \Delta t, d_0, f\}$.

\subsection{Ferromagnetic case}
\label{sec:numferro}

\begin{figure}[t]
\epsfxsize=9cm
\epsfbox{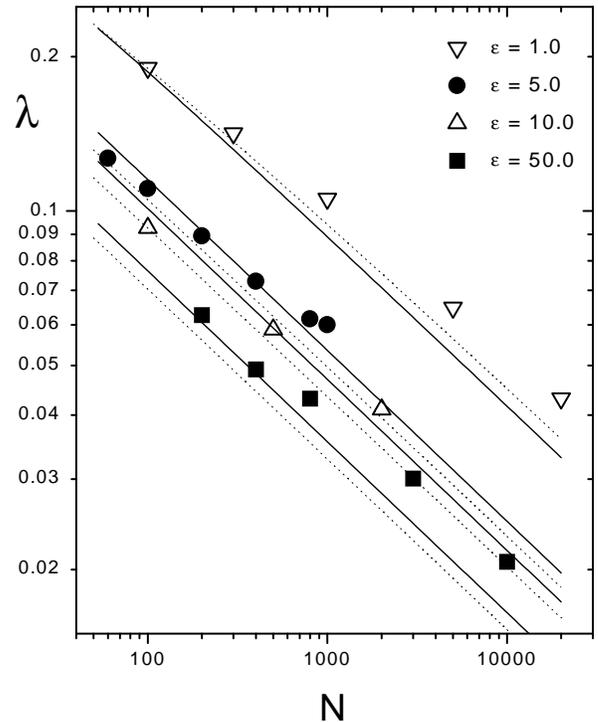}
\vspace{2pc}
\caption{  Largest Lyapunov exponent of the mean-field XY-Hamiltonian ($J=1$)
           as a function of system size $N$, for some selected 
           energies $\varepsilon$.
           From top to bottom, $\varepsilon=1.0,5.0,10.0,50.0$, respectively.
           Symbols correspond
           to numerical simulations of Hamilton's equations. 
           Full lines are our theoretical results; dotted lines correspond to
           the prediction of the ``geometric method" (see text).}
\label{fig:liapferro}
\end{figure}

Figure~\ref{fig:liapferro} shows the largest Lyapunov exponent of 
the ferromagnetic HMF as a function of system size $N$, for some selected 
energies $\varepsilon$. 
The symbols correspond to the simulations of 
Refs.~\cite{anteneodo98} ($\varepsilon=5.0$) and \cite{latora98} ($\varepsilon=1.0,10.0,50.0$). 
We have considered large particle numbers $(N \ge 100)$
and energies well above the transition ($\varepsilon>1.0$) to ensure 
(i) the validity of the approximations invoked in calculating the averages of
Sect.~\ref{sec:XY}, and also (ii) to guarantee that we are in a disordered, 
quasi-ballistic regime. Full lines correspond to the theoretical Lyapunov exponent
obtained by diagonalizing the $3\times 3$ matrix of Sect.~(\ref{sec:theory}), but  
we could as well have used the asymptotic expression of Eq.~(\ref{lambda-asymp}) 
(minor differences would only be visible in the case $\varepsilon=1$). 
Dotted lines correspond to the geometric prediction, as calculated by Firpo,
i.e., Eqs.~(1-3) and (21-22) of Ref.~\cite{firpo98}.
An inspection of Fig.~\ref{fig:liapferro} allows us to verify that there is a 
satisfactory agreement between theory and simulation, especially for $N \le 500$.
For larger systems numerical data deviate upward from theory.
(Nobre and Tsallis also observed deviations from the $N^{-1/3}$ law in the three-dimensional
version of the XY-Hamiltonian \cite{nobre03}.) 
This is the opposite to natural expectation 
``the larger the system, the better the approximations".  
We believe that the reason for this deviation may be that the system has not reached 
microcanonical equilibrium -- as far as the Lyapunov exponent is concerned. 

It is well known that the HMF relaxes very slowly, e.g., 
Latora, Rapisarda, and Ruffo \cite{latora99} noticed that for $\varepsilon =1$, 
the system may get trapped in quasi-equilibrium states similar to those observed 
below the transition.
A very slow relaxation led the same authors to conclude that dynamics is ballistic
for $\varepsilon=5.0$, $N=1000$, $t<10^5$ \cite{latora99b}.
If slow relaxation is ruling, the numerical
simulations are likely to provide ``quasi-equilibrium" Lyapunov exponents, rather than the 
microcanonical ones. Actually, as energies and/or system sizes are increased, the system 
becomes progressively more integrable and deviations from microcanonical predictions 
will be necessarily observed.

\begin{figure}[ht]
\epsfxsize=8cm
\epsfbox{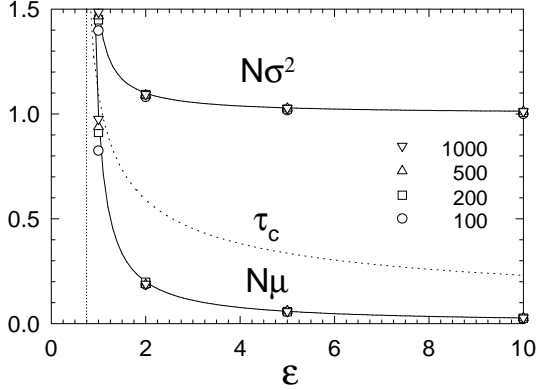}
\vspace{-4pc}
\caption{Numerical (symbols) and analytical (lines) values of $\mu$, $\sigma^2$,
and $\tau_c$ for
the HMF model ($J=1$) as functions of energy 
$\varepsilon$ and system size $N=100,200,500,1000$. 
Error bars are of the order of the symbol size, or smaller.}
\label{fig:musi}
\end{figure}

At this point it is desirable to verify if the canonical calculations of 
$\mu$, $\sigma^2$, and $\tau^{(k)}$, reproduce the corresponding dynamical averages.
In Fig.~\ref{fig:musi} we show numerical values for $\mu$ and $\sigma^2$, obtained through 
averaging over 1000 initial conditions, and in the time window $[5000,10000]$.  
One sees that $\sigma^2$ is very close to the theoretical value, within a few percent, but
the results for $\mu$ show larger deviations. 
We can understand why $\sigma^2$ is so close to its expectation value and $\mu$ is not:
The initial value of $\sigma^2$ is already very close to the canonical value, within a few
percent for the cases considered. On the contrary, the initial value for $\mu$ is zero, far
from its equilibrium value; the slow growth of $\mu$ is one of the aspects of the slow relaxation
to equilibrium.
In any case we verified (graphics not shown) that the errors in $\mu$ and $\sigma^2$
do not account for the deviations of Lyapunov exponents, i.e., feeding the theory  
with the numerical $\mu$ and $\sigma^2$ does not improve the agreement between theoretical 
and simulated $\lambda$'s. 

\begin{figure}[ht]
\epsfxsize=8cm
\epsfbox{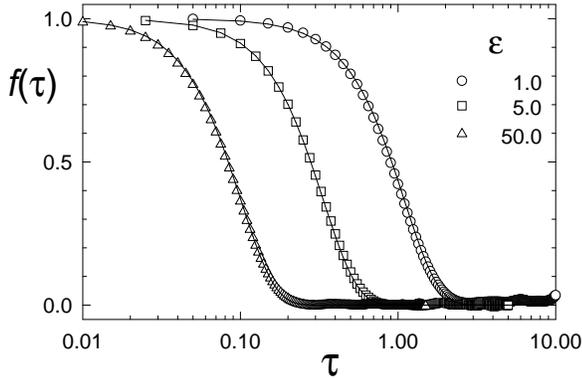}
\vspace{-4pc}
\caption{Correlation functions for $N=200$ and several energies,
from right to left, $\varepsilon=1.0,5.0,50.0$ ($J=1$). Symbols
correspond to simulations with $t_{\rm eq}=10000$ and averaging over 
100 initial conditions.
The lines represent our theoretical prediction (Gaussians).}
\label{fig:corr}
\end{figure}

In order to discuss the correlation times, let us now look at the correlation functions 
of Fig.~\ref{fig:corr}.
Theory and simulations agree almost perfectly in the central part of the distributions.
There are long tails, which cannot be appreciated in the figure but might
be responsible for larger-than-predicted correlation times $\tau^{(k)}$.
We recall that the quantities $\tau^{(k)}$ are moments of the correlation function 
$f(\tau)$, and as such, very sensitive to the precise shape of the tails 
We verified that a pure numerical evaluation of $\tau^{(k)}$ fails due to poor
convergence of the corresponding integrals -- this effect is stronger for $k=1,2$.
So, at present we cannot say with certainty if the theoretical estimates for $\tau^{(k)}$,
which assume a Gaussian correlation function, agree with the corresponding dynamical ones.
To resolve this issue one should make a very careful study of the tails of the correlation
functions. Though desirable such analysis exceeds the scope of this paper. 
(Interesting information about correlation functions, and, in general,
 about the geometric method, can be found in Refs.~\cite{kandrup00,terzic03}.) 

In passing, let us comment that, for the theory to work, $f(\tau)$ must decay fast enough, 
e.g., a correlation function like $\Gamma(\tau) = \sin \omega t /\omega t$ \cite{casetti96,kandrup00}
would explode the second cumulant.

To conclude the analysis of the ferromagnetic case,
a possible difference between $\lambda$ and $\lambda^\star$ 
[Eqs.~(\ref{liapunovnum}) and (\ref{liapunovnum2})] cannot 
be the explanation for the observed deviations between theory and simulations because 
it is expected that $\lambda < \lambda^\star$ (see below, and Ref.~\cite{benzi85}).

\subsection{Antiferromagnetic case}
\label{sec:numanti}

\begin{figure}[ht]
\epsfxsize=8cm
\epsfbox{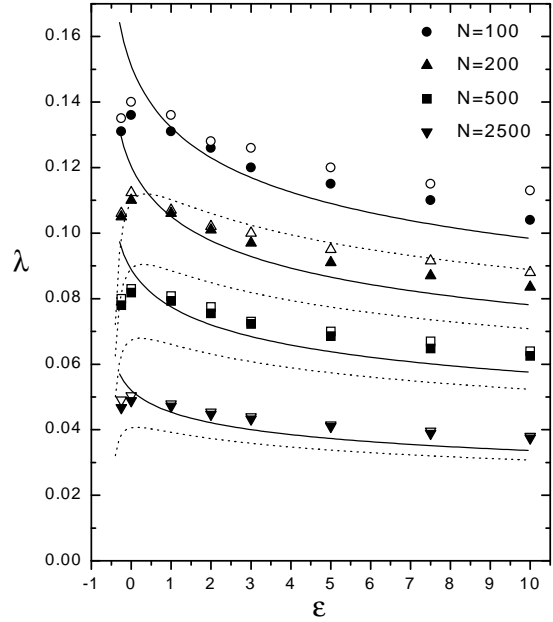}
\vspace{1pc}
\caption{Lyapunov exponent of the anti-ferromagnetic mean-field Hamiltonian 
         model as a function of energy $\varepsilon$. 
         Symbols are the result of simulations using Benettin's algorithm.
         Open symbols correspond to the usual Lyapunov exponent $\lambda$ 
         and hollow 
         symbols to $\lambda^\star$ (see text).
         We used $t_{\rm eq}=1000$ and twin trajectories were propagated during 
         $t_{\rm prop}=3000$. 
         Full lines correspond to our theory, dotted ones to the 
         geometric method.}
\label{fig:ateovsnum}
\end{figure}

Figure~\ref{fig:ateovsnum} shows the results of numerical simulations
for $\lambda$ and $\lambda^\star$. 
(The numerical results for $\lambda$, in the case $N=100$, are consistent 
 with those obtained in Ref.~\cite{latora00}.)
Though not shown, as temperature goes to zero ($\varepsilon \to -1/2$), the numerical 
$\lambda$'s go to zero too, the empirical scaling law being $\lambda \propto T^{1/2}$ \cite{latora00}.
Notice that the relative differences between $\lambda$ and $\lambda^\star$ are indeed small, and 
decrease with increasing $N$. We have not analyzed the behavior of $\lambda^\star$ at very low 
temperatures; in particular, we cannot say whether the scaling 
law $\lambda^\star \propto T^{1/2}$ is verified or not. 

\begin{figure}[ht]
\epsfxsize=8cm
\epsfbox{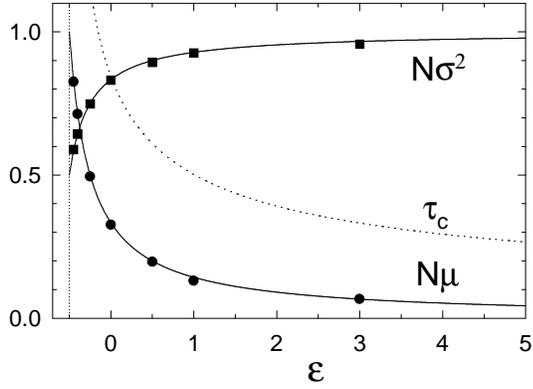}
\vspace{-4pc}
\caption{Numerical and analytical values of $\mu$ and $\sigma^2$ for
the mean-field XY-Hamiltonian model as functions of energy 
$\varepsilon$ for $N=100$ ($J=-1$). 
The results shown are time averages in the window $t \in [5000,10000]$,
and over 100 initial conditions.
Lines correspond to the corresponding canonical averages.}
\label{fig:amusi}
\end{figure}

For high energies, i.e., $\varepsilon > 1$, our theoretical predictions 
agree reasonably well with simulations. However, when the energy tends
to its ground-state value, $\varepsilon \to -1/2$, our theoretical Lyapunov 
exponent diverges dramatically from experiments. These behaviors can be
easily understood by looking at the energy dependence of the perturbative
parameter $\sigma \tau$. Figure~(\ref{fig:amusi}) shows that, for a fixed system size,
$\sigma^2$ remains bounded for all energies, growing by a factor of two as $T$ goes 
from zero to infinity. On the other side, $\tau$ scales with temperature like 
$1/\sqrt{T}$, and then, $\sigma \tau \propto 1/\sqrt{T}$. This means that the theory must
fail as $T \to 0$. The divergence of the Kubo number at $T=0$ is a manifestation of 
the complexity of the dynamics close to the antiferromagnetic ground state 
\cite{reviewHMF,bicluster}. Notwithstanding the overall agreement at high temperatures,
we must point out a non-resolved source of annoyance. At a fixed energy the Kubo parameter scales
with the system size like $\sigma \tau \propto 1/\sqrt{N}$. Consequently the relative
error of the theoretical prediction should decrease 
with increasing $N$, like $1/\sqrt{N}$. However, e.g.,
if we take $\varepsilon=7.5$, and compare theory versus $\lambda^\star$, we see that 
the relative error is about 11\%, 
insensitive to variations
of $N$ from 100 to 2500.

\begin{figure}[ht]
\epsfxsize=8cm
\epsfbox{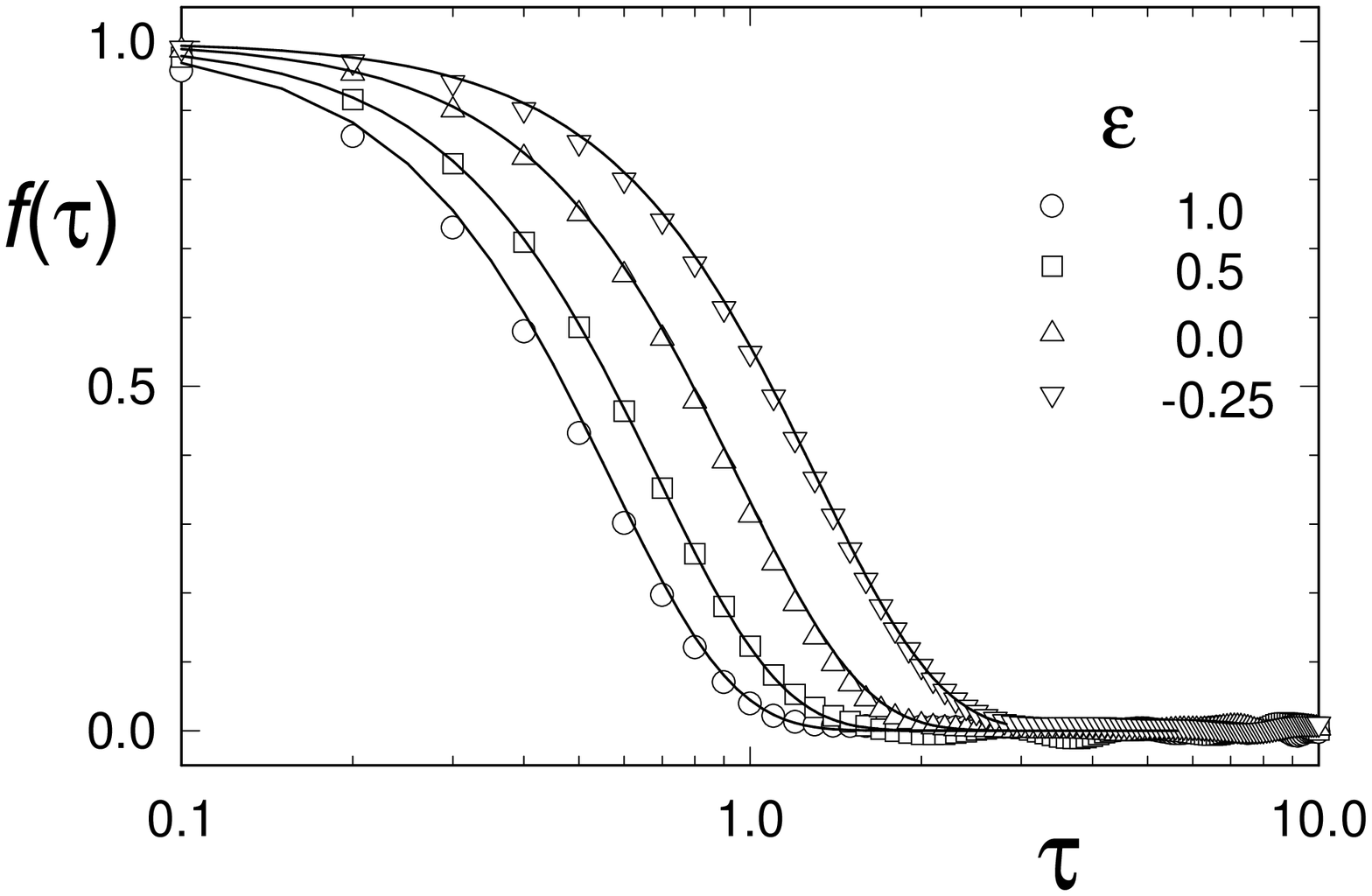}
\caption{Correlation functions for $N=100$ and several energies ($J=-1$).
         Averaged over 100 initial conditions. Equilibration time, $t_{\rm eq}=10000$.
         Solid lines are our analytical results.}
\label{fig:acorr}
\end{figure}

For the sake of completeness, we exhibit in Figs.~(\ref{fig:amusi}) and (\ref{fig:acorr})
the comparison of canonical and dynamical averages for $\mu$, $\sigma$, and $f(\tau)$:
No significant differences can be seen. (Again, the tails of the correlation functions
may deserve a deeper analysis.)

Figure~(\ref{fig:ateovsnum}) also displays the predictions of the geometric 
method \cite{firpo98}. In the high-temperature regime, the relative error of the
geometric prediction is somewhat larger than ours. At low temperatures, the 
geometric method predicts a qualitatively correct behavior, $\lambda \to 0$, but 
with the wrong scaling law:  $\lambda \propto T^2$ \cite{firpo98} instead of 
$\lambda \propto T^{1/2}$ \cite{latora00}.

\section{Summary and concluding remarks}
\label{sec:conclusions}

In a previous paper \cite{vallejos02} we proposed a theoretical approach to the 
determination of the largest Lyapunov exponent of many-particle Hamiltonian systems.
Despite several crude approximations that had to be made, it was not completely clear
whether the resulting perturbative/mean-field script could be carried out for a specific 
system, one of the obstacles being the analytical determination of the correlation functions.

Now we have verified that the stochastic recipe can indeed be executed and works 
satisfactorily in the quasi-ballistic regimes of the mean-field XY-Hamiltonians.
These systems are especially useful for testing the theory because they make the 
``mean-field" diagonalization exact. However, three additional sources of error 
have to be considered. 
First of all, we made the simplification of taking the logarithm out of the average 
[Eq.~(\ref{defliapunov2})]. 
That is, using a spin-glass analogy, we are estimating an ``annealed" Lyapunov exponent 
$\lambda^\star$ instead of the usual ``quenched" $\lambda$ \cite{tanase02}. 
Quenched averages can be calculated, as shown by T\u{a}nase-Nicola and Kurchan \cite{tanase02}, 
but they require more sophisticated tools, like the replica trick or the supersymmetric approach.
In the cases considered in this paper we verified that the difference between 
both Lyapunov exponents is very small. Then it was not necessary to go beyond the annealed 
averaging. 

The second approximation, and most important one, is the truncation of the cumulant expansion at 
the second order.
By doing so, we introduce a relative error of the order of the Kubo number $\sigma \tau_c$.
This explains the failure of the theory in the low-temperature regime of the 
antiferromagnetic XY-Hamiltonian, given that the amplitude of the fluctuations $\sigma$ 
remains bounded but the thermal time $\tau_c$ diverges.
It is not clear at present if
the theory can be improved to account for this regime. 
Further studies are necessary. 
In particular we have to understand what kind of dynamical correlations develop as the system
approaches the ground state. 
Another problem that requires an answer: We have not observed
that the agreement between theory and simulations betters when the size of the system
is increased, for fixed energy. This is in contradiction with the Kubo number decreasing like
$1/\sqrt{N}$.

The difficulties of the preceding paragraph may be related to the third approximation, which 
concerns the interaction representation we have used. In order to isolate the fluctuations,
one should work in the representation associated to the average Hessian. Instead, we have chosen 
the free-rotator representation. That is, in constructing the ``free propagator", we neglected 
the average Hessian of the interaction potential. As long as the motion is quasi-ballistic, this 
approximation seems justified. It will be certainly wrong, e.g., in the low-temperature phase of 
the ferromagnetic XY, where the quadratic part of the interaction is really important.
We are currently working on the implementation of the best interaction representation for the 
infinite-range XY models. The results will be presented in a forthcoming paper.

The mean-field diagonalization is exact for the systems considered here. In other cases it 
represents just a truncation of the basis for diagonalizing $\hat{\Lambda}$ [Eq.~(\ref{expansion})]. 
Notice, however, that we are dealing with a hermitian problem, i.e., finding the largest 
eigenvalue of $\langle \xi \xi^T \rangle(t)$ [Eq.~(\ref{solution})].
Truncation of the basis will produce a lower bound to the largest Lyapunov exponent (provided
that $\hat{\Lambda}$ is calculated accurately). 
So, this problem is analogous to finding the ground-state energy of a quantum Hamiltonian.
Any small subspace can be considered, the choice being guided by the special characteristics 
of the system under study. 

Like its quantum analogue, the Lyapunov problem seems reluctant to admit simple general solutions.
Each class of systems may require special consideration.


\section*{Acknowledgements}


We are grateful to F. V. Roig and C. Tsallis for fruitful discussions.
We acknowledge Brazilian Agencies CNPq, FAPERJ and PRONEX for financial
support.


%
%

\newpage
%
%
%
%
%

%
%

\newpage
%
%
%
%
%

\end{multicols}


\begin{thebibliography}{99}


\bibitem{vallejos02} 
R. O. Vallejos and C. Anteneodo, Phys. Rev. E {\bf 66}, 021110 (2002).

\bibitem{geometric} 
L. Casetti, R. Livi and M. Pettini, Phys. Rev. Lett. {\bf 74}, 375 (1995);
L. Casetti, M. Pettini and E. G. D. Cohen, Phys. Rep. {\bf 337}, 238 (2000).

\bibitem{casetti96}
L. Casetti, C. Clementi and M. Pettini, Phys. Rev. E {\bf 54}, 5969 (1996).

\bibitem{barnett96} 
D. M. Barnett, T. Tajima, K. Nishihara, Y. Ueshima, and H. Furukawa, 
Phys. Rev. Lett. {\bf 76}, 1812 (1996) 
and 
Phys. Rev. E {\bf 55}, 3439 (1997);
D. M. Barnett and T. Tajima, Phys. Rev. E {\bf 54}, 6084 (1996).
See also
A. Torcini, Ch. Dellago and H. A. Posch, Phys. Rev.  Lett. {\bf 83}, 2676 (1999); 
D. M. Barnett et al., Phys. Rev. Lett. {\bf 83}, 2677 (1999).

\bibitem{benettin76}  
G. Benettin, L. Galgani, and J.-M. Strelcyn, Phys. Rev. A. {\bf 14}, 2338 (1976).

\bibitem{vankampen} 
N. G. van Kampen, Phys. Rep. {\bf 24}, 171 (1976);
{\it Stochastic Processes in Physics and Chemistry} (North-Holland, Amsterdam, 1981).

\bibitem{reviewHMF} 
T. Dauxois, V. Latora, A. Rapisarda, S. Ruffo, and A. Torcini,
e-print cond-mat/0208456.

\bibitem{antoni95}  
M. Antoni and S. Ruffo, Phys. Rev. E {\bf 52}, 2361 (1995).

\bibitem{metastable} 
V. Latora, A. Rapisarda, and S. Ruffo, Physica A {\bf 280}, 81 (2000);
V. Latora, A. Rapisarda, and C. Tsallis, Phys. Rev. E {\bf 64}, 056134 (2001); 
V. Latora and A. Rapisarda, Chaos, Solitons and Fractals {\bf 13}, 401 (2002); 
A. Giansanti, D. Moroni, and A. Campa, {\it ibid.}, p. 407; 
V. Latora, A. Rapisarda, and C. Tsallis, Physica A {\bf 305}, 129 (2002);
A. Pluchino, V. Latora, and A. Rapisarda, e-print cond-mat/0303081.

\bibitem{tsallis}
G. Kaniadakis, M. Lissia, and A. Rapisarda, eds., 
Non Extensive Thermodynamics and Physical Applications, 
Physica A {\bf 305} (Elsevier, Amsterdam, 2002); 
%
M. Gell-Mann and C. Tsallis, eds., Nonextensive Entropy - Interdisciplinary 
Applications (Oxford University Press, Oxford, 2003), in press; 
%
H. L. Swinney and C. Tsallis, eds., Anomalous Distributions, Nonlinear Dynamics, 
and Nonextensivity, Physica D (Elsevier, Amsterdam, 2003), in press. 
%
A regularly updated bibliography on the subject can be found 
at http://tsallis.cat.cbpf.br/biblio.htm.


\bibitem{bicluster}
T. Dauxois, P. Holdsworth, and S. Ruffo, Eur. Phys. J. B. {\bf 16}, 659 (2000);
J. Barr\'e, T. Dauxois, and S. Ruffo, Physica A {\bf 295}, 254 (2001);
J. Barr\'e, F. Bouchet, T. Dauxois, and S. Ruffo, Eur. Phys. J. B. {\bf 29}, 577 (2000).

\bibitem{antoni02}
M. Antoni, H. Hinrichsen, and S. Ruffo, 
Chaos, Solitons and Fractals {\bf 13}, 393 (2002).

\bibitem{velazquez03} 
L. Velazquez, R. Sospedra, J.C. Castro, and F. Guzm\'an,
e-print cond-mat/0302456.

\bibitem{latora99}  
V. Latora, A. Rapisarda, and S. Ruffo, Physica D {\bf 131}, 38 (1999).

\bibitem{yamaguchi96}    
Y. Y. Yamaguchi, Prog. Theo. Phys. Supp. {\bf 95}, 717 (1996).

\bibitem{latora00}  
V. Latora, A. Rapisarda and S. Ruffo, Prog. Theo. Phys. Supp. {\bf 139}, 204 (2000).

\bibitem{firpo98}
M.-C. Firpo, Phys. Rev. E {\bf 57}, 6599 (1998).

\bibitem{anteneodo01}  
C. Anteneodo and R. O. Vallejos, Phys. Rev. E {\bf 65}, 016210 (2001).

\bibitem{tanase02} 
S. T\u{a}nase-Nicola and J. Kurchan, e-print cond-mat/0210380.

\bibitem{hughes95}
B. D. Hughes, {\em Random Walks and Random Environments}, Vol. 1 (Clarendon Press, New York, 1995).

\bibitem{wolfram91}
S. Wolfram, {\em Mathematica: A System for Doing Mathematics by Computer}, 2nd. Ed.
(Addison-Wesley, Reading, 1991). 
    
\bibitem{latora98}  
V. Latora, A. Rapisarda and S. Ruffo, Phys. Rev. Lett. {\bf 80}, 692 (1998).

\bibitem{anteneodo98} 
C. Anteneodo and C. Tsallis, Phys. Rev. Lett. {\bf 80}, 5313 (1998).

\bibitem{yoshida90}
H. Yoshida, Phys. Lett. A {\bf 150}, 262 (1990).

\bibitem{nobre03}
F. D. Nobre, and C. Tsallis, e-print cond-mat/0301492.

\bibitem{latora99b} 
V. Latora, A. Rapisarda, and S. Ruffo, Phys. Rev. Lett. {\bf 83}, 2104 (1999).

\bibitem{kandrup00}
H. E. Kandrup, I. V. Sideris, and C. L. Bohn, Phys. Rev. E {\bf 65}, 016214 (2001).

\bibitem{terzic03}
B. Terzi\'c and H. E. Kandrup, Phys. Lett. A {\bf 311}, 165 (2003).

\bibitem{benzi85}
R. Benzi, G. Paladin, G. Parisi, and A. Vulpiani, J. Phys. A {\bf 18}, 2157 (1985);
H. Schomerus and M. Titov, Phys. Rev. E {\bf 66}, 066207 (2002).


\end{thebibliography}
\end{document}